\documentclass[aps,prd,10pt,twocolumn,superscriptaddress,nofootinbib,showkeys,showpacs,altaffilletter]{revtex4-1}

\usepackage{graphicx}
\usepackage{dcolumn}
\usepackage{amssymb}
\usepackage{amsmath}
\usepackage{amsfonts}
\usepackage{amsbsy}
\usepackage{color}
\usepackage{rotating}
\usepackage[english]{babel}
\usepackage{multirow}

\newcommand{\be}{\begin{equation}}
\newcommand{\ee}{\end{equation}}
\newcommand{\bea}{\begin{eqnarray}}
\newcommand{\eea}{\end{eqnarray}}

\begin{document}

\title{Probing the constancy of the speed of light with future galaxy survey: the case of SKA and \textit{Euclid}}

\date{\today}

\author{Vincenzo Salzano}
\email{enzo.salzano@wmf.univ.szczecin.pl}
\affiliation{Institute of Physics, University of Szczecin, Wielkopolska 15, 70-451 Szczecin, Poland}
\author{Mariusz P. D\c{a}browski}
\affiliation{Institute of Physics, University of Szczecin, Wielkopolska 15, 70-451 Szczecin, Poland}
\affiliation{Copernicus Center for Interdisciplinary Studies, S{\l}awkowska 17, 31-016 Krak{\'o}w, Poland}
\affiliation{National Centre for Nuclear Research, Andrzeja So{\l}tana 7, 05-400 Otwock, Poland}
\author{Ruth Lazkoz}
\affiliation{Fisika Teorikoaren eta Zientziaren Historia Saila, Zientzia eta
Teknologia Fakultatea, \\ Euskal Herriko Unibertsitatea, 644 Posta Kutxatila,
48080 Bilbao, Spain}


\begin{abstract}
In \citep{VSL_us} a new method to measure the speed of light through Baryon Acoustic Oscillations (BAO) was introduced. Here, we describe in much more detail the theoretical basis of that method, its implementation, and give some newly updated results about its application to forecast data. In particular, we will show that SKA will be able to detect a $1\%$ variation (if any) in the speed of light at $3\sigma$ level. Smaller signals will be hardly detected by already-planned future galaxy surveys, but we give indications about what sensitivity requirements should a survey fulfill in order to be successful.
\end{abstract}

\keywords{Cosmology, Baryon Acoustic Oscillations, speed of light}

\pacs{$98.80-k,98.80.Es,98.80.Cq, 04.50.Kd$}


\maketitle

\section{Introduction}

The speed of light is one of the most fundamental constants of nature playing a significant role in basic physical laws such as the Maxwell equations, special and general relativity equations, atomic and particle physics equations and many others. In other words, it influences the vast areas of physics which deal with both microscopic and astronomical scales. Because of its crucial meaning, the speed of light was officially announced in 1983 to take a fixed value by BIPM (Bureau International des Poids et Mesures) \citep{velocity_c_1}. Lots of measurements of the speed of light have been performed, beginning from the famous, though inaccurate, measurement by R$\o$mer and Huygens in 1675, following through Bradley, Fizeau, and Michaelson, and ending in a laser interferometric measurement by \citep{velocity_c_2}.  However, in view of the contemporary theories of physics such as multidimensional theories of gravity within the framework of superstring and brane theories (see e.g., \citep{polchin}), some physicists argue that the values of physical constants like the gravitational constant $G$, the fine structure constant $\alpha$, the electron-to-proton mass ratio $m_e/m_p$, and the speed of light $c$ may become dynamical (represented by some scalar fields for example), and so they can evolve in time and space (for a review of such a variation, see \citep{Uzan2011}).

The option which has probably the strongest impact on the whole physics is the variability of $c$. Such an idea even dates back to Einstein himself \citep{VSL_old}, but it has attracted more interest recently due to the fact that it can provide an alternative solution of the classic problems of non-inflationary cosmology such as the horizon and the flatness problems. Having such advantages, the theories of varying speed of light -- in short VSL theories -- are often regarded as controversial because they are usually not formulated in the proper framework of dynamical scalar field theory  \citep{VSL_controversy}, allowing a special choice of the frame in which the speed of light is constant \citep{VSL_theory,Magueijo2003}, though there are some attempts at a more proper formulation \citep{new_Moffat}. In particular, a comparison of those theories with experimental data seems to be still missing \citep{VSL_observations}. In this paper, following our brief previous study \citep{VSL_us}, we will try to put constraints on some varying speed of light theories by using future galaxy surveys such as the Square Kilometer Array (SKA) \citep{SKA_site}, \textit{Euclid} \citep{EUCLID_1,EUCLID_2}, and the Wide-Field Infrared Survey Telescope \textit{WFIRST-2.4} \citep{WFIRST}, showing how the huge number of galaxies which will be collected can be used as a probe for the constancy of $c$.

One of the main signals that can be detected by a galaxy survey is related to the Baryon Acoustic Oscillations (BAO) \citep{BAO_review}. Well theoretically established since $1970$ \citep{BAO_origins}, they have been recognized as one of the most useful and promising probes for studying dark energy and cosmology only relatively recently \citep{BAO_recog_0,BAO_recog_1}. At the present stage, even if they are very helpful, they are still far from being at their best use; that is the reason why they are among the main objectives of many important on-going and future earth-based and spatial surveys, such as SKA, \textit{Euclid}, \textit{WFIRST-2.4}, the Baryon Oscillation Spectroscopic Survey (BOSS) \citep{BOSS_0}, the Extended BOSS survey (eBOS)S \citep{eBOSS}, the Dark Energy Spectroscopic Instrument (DESI) \citep{DESI} and the Hobby-Eberly Telescope Dark Energy Experiment (HETDEX) \citep{HETDEX}. Basically, BAO observational outputs are measurements of the sound horizon at late times, as it is imprinted in the clustering of large scale structure. 
It is generally considered as a standard ruler, i.e., an ``object'' whose size is constant in time and that can be used as a stick to calibrate/measure cosmological distances (for problems and alternatives, see \citep{Anselmi}). Its magnitude can be exactly calculated from the theory and it is approximately equal to $150$ Mpc in physical units. This is just the value which can be measured, with the best precision possible, from the Cosmic Microwave Background (CMB) observations. The latest data release from \textit{Planck} \citep{Planck_cosmo} gives us a value of $r_{s}(z_{rec}) = 144.81 \pm 0.24$ Mpc for the baseline model, exhibiting a very weak cosmological model dependence. Given the strong correlation between photons (measured by CMB) and baryons, such a distance should be imprinted in the large scale structure; and so it is. Measuring the distribution of galaxies in space, as well as their redshifts, and analyzing their correlation function, it is possible to observe the typical correlation length which, as expressed in comoving units, corresponds exactly to the sound horizon. Being more precise, it corresponds to the sound horizon not at recombination, but rather at a later epoch defined as ``dragging redshift'' \citep{BAO_recog_0}, at $z_{drag} \approx 1060$. Of course, galaxy distribution is three-dimensional, and the sound horizon should be measured in three different directions: two are on the projected sky, and one is in the radial direction. The former are said to be the tangential modes, the latter the radial. They can be defined as
$$
y_{t}(z) = \frac{D_{A}(z)}{r_{s}(z_{rec})} \quad \mathrm{and} \quad y_{r}(z) = \frac{c}{H(z)r_{s}(z_{rec})} \, ,
$$
where $c$ is the speed of light, $z$ is the cosmological redshift, $D_{A}$ is the angular diameter distance, $H$ is the Hubble function, and $r_{s}(z_{dec})$ is the sound horizon, evaluated at recombination (or dragging epoch).

Nowadays, we do not have yet such a good signal in order to have accurate measurements for $y_t$ and $y_r$ separately. We have good measurements of quantities combining $D_A$ and $H$, as, for example, the average distance
\bea
\label{av_dist}
D_{V} = \left[(1+z)^{2} c \; z \frac{D_{A}^{2}}{H} \right]^{1/3},
\eea
or the Alcock-Paczy$\acute{\mathrm{n}}$ski distortion parameter
\bea
\label{dist_par}
F = (1+z) D_{A} \; \frac{H}{c}.
\eea
There are some trials to obtain independent information for $D_A$ and $H$ \citep{WiggleZ,BOSS}, but they are not yet fully competitive. With future surveys, with a larger number of galaxies available, this will be possible eventually, and it will reveal as a necessary requirement for our method to be applied.

The same galaxies used for detecting BAO, or, at least, a fraction of them, can also be used as {\it cosmic chronometers}. The seminal idea of cosmic chronometers was first described in \citep{chronometers_0}, and then progressively extended and used for cosmological analysis in \citep{chronometers_1,chronometers_2}. It is based on the \textit{differential age method}: the key is to find a ``cosmological clock'', able to return the variation of Universe age with redshift. If one has this clock, then, one simply has to measure the age difference $\Delta t$ between to redshifts separated by $\Delta z$, and calculate from these the derivative $d z / d t \approx \Delta z / \Delta t$. This latter quantity then would be directly related to the Hubble function, defined as
$$
H(z) = -\frac{1}{1+z} \frac{dz}{dt} \; .
$$
If such method were possible, we would have a measurement of the Hubble function free from any assumption on the nature of the metric, which normally affects, for example, the definition of cosmological distances. It was proposed in \citep{chronometers_0} that the role of such clocks could be played by passively-evolving early-type galaxies (ETG). It has also been shown how to use them, and what order of constraints should be expected from them. Since then, a lot of work and improvements have been made: now we have better stellar population models; we have a much larger number of observations (see e.g. \citep{chronometers_1} in which $\approx 10^4$ galaxies were analyzed) and very deep in redshift (up to $z \sim 2$); as well as more precise tools to calibrate the clocks (the $4000$ {\AA} break in ETG spectra). And this scenario can still be improved using future galaxy surveys in the optical, as \textit{Euclid} and \textit{WFIRST-2.4}, which should observe at least ten times more galaxies compared to the present (and ETG, eventually).

The paper is organized as follows: in Sec.~\ref{sec:Theory} we will describe the theoretical background underlying the proposed method; in Sec.~\ref{sec:Method} we will describe in details all the steps involved in the building of our method; and finally in Sec.~\ref{sec:Results} we will apply our method to some particular cases and will discuss the results.

\section{Theoretical basis}
\label{sec:Theory}

The possibility to constrain VSL theories from large scale structure is strictly related to the definition of one of the quantities that can be measured in a galaxy (BAO) survey, i.e., the angular diameter distance, $D_{A}(z)$. This distance is defined as
\begin{equation}
D_{A}(z) = \frac{1}{1+z}\int^{z}_{0} \frac{c_{0}}{H(z')} \, d z' \; ,
\end{equation}
where $c_{0}$ is the speed of light. From now on, we will assume the convention to define $c_{0} \equiv 299792.458$ km s$^{-1}$ as the value of the speed of light \citep{velocity_c_1}. This is, of course, assumed to be constant in a standard scenario (and in most of the physics, nowadays); while, in a VSL theory, it is equal to the speed of light evaluated \textit{here and now}.

A very well known, and somewhat counterintuitive property of $D_{A}(z)$, is that it rises up to a maximum at some redshift, which we will call $z_{M}$, and then starts to decline \citep{Weinberg}. Starting from its definition, an equivalent way to set the problem is to say that the angular size of a given escaping object diminishes while it is going farther from us up to some point, where it reaches a minimum, before it starts rising again. Both pictures basically tell us that early times objects (or, at least, older than some redshift $z_{M}$) look closer than late times ones. The explanation behind this peculiar behavior is a mix of geometric facts (curvature, non-Euclidean space) and the dynamical history of our Universe \citep{DAprop}.

The exact location of the maximum, i.e. the redshift $z_{M}$, depends on the cosmological model, which enters the definition of $H(z)$. In order to have a general idea of the range possibly covered by $z_{M}$, and compatible with the most updated set of cosmological probes available, we have considered the CPL \citep{CPL} $w+w_{a}$ \texttt{plikHM}$\_$\texttt{TTTEEE}$\_$\texttt{lowTEB}$\_$\texttt{BAO}$\_$\texttt{post}$\_$\texttt{lensing}
bestfit from the \textit{Planck 2015} release \citep{Planck_Archive}. We have considered a total of $10^4$ cosmological models, derived from varying the cosmological parameters consistently with the $1\sigma$ confidence intervals defined for the previous parametrization. Of course, the $w+w_{a}$ is only one of the many dark energy models available, but it is somewhat used as a ``reference'' model in the literature. Moreover, the large errors on its parameters, in particular on the dynamical dark energy EoS parameter $w_{a}$, make us confident on having explored a very large set of cosmological scenarios compatible with observational data, thus making our estimation for the range of $z_{M}$ highly conservative. We have checked that $z_{M}$ lies in the range $[1.4,1.75]$ for more than $99\%$ of $10^{4}$ random cosmological models chosen as described above. This is a quite narrow redshift range and, what is very interesting, will be covered by many surveys in the future (SKA, \textit{Euclid} and \textit{WFirst2.4}), so that we will have good quality data in such range.

Given the dependence of $z_M$ on the cosmological model, one could think about using it as a further tool to constrain, for example, dark energy properties, in addition to the most used probes in cosmology. Unfortunately, the large degeneracy between the cosmological parameters in such a narrow range, as it was shown in the previous case for the CPL parametrization, makes $z_{M}$ of no real use in such case \citep{DAcosmo}. But we have found a different and a very interesting way for which $z_M$ can be usefully used to explore the nature of our Universe.

In fact, a very interesting relation (in the context of testing the validity VSL) exists between the angular diameter distance and the maximum redshift, which is easy to derive and intrinsic to its definition: the mathematical condition for the maximum of a function  is that the derivative with respect to a variable vanishes. In the case of $D_A$, we have that the condition $\partial D_{A}(z) / \partial z = 0$, when evaluated at $z_{M}$, corresponds to the relation:
\begin{equation}\label{eq:relation}
D_{A}(z_{M}) = \frac{c_{0}}{H(z_{M})} \quad \Rightarrow \quad D_{A}(z_{M})H(z_{M}) = c_{0}\, ,
\end{equation}
i.e., the multiplication of the angular diameter distance and the Hubble function, both evaluated at the maximum redshift, will give the value of the speed of light. It is worth to underline here that such relation is not fully model independent, but it is based on two hypothesis: a Friedmann-Robertson-Walker (FRW) metric of the background; and no spatial curvature, i.e. $k=0$. While the former is quite general and it is used as an assumption in most of the models on the market (despite some non-Friedmannian models are theoretically studied anyway \citep{inhomogeneous}), the latter has to be proven not to be invalidating our results. At least, it can be shown that even in the case of $k \neq 0$, Eq.~(\ref{eq:relation}) has still some validity. Allowing the curvature to be different from zero, the angular diameter distance is defined as:
\begin{equation}\label{eq:DA_k}
D_{A}(z) = \begin{cases} \frac{D_{H}}{\sqrt{\Omega_{k}}(1+z)} \sinh \left( \sqrt{\Omega_{k}} \frac{D_{C}(z)}{D_{H}} \right) &\mbox{for } \Omega_{k} > 0\\
\frac{D_{C}(z)}{1+z} &\mbox{for } \Omega_{k} = 0 \\
\frac{D_{H}}{\sqrt{|\Omega_{k}|}(1+z)} \sin \left( \sqrt{|\Omega_{k}|} \frac{D_{C}(z)}{D_{H}} \right) &\mbox{for } \Omega_{k} < 0 \, ,
\end{cases}
\end{equation}
where $D_{H} = c_{0} / H_{0}$ is the Hubble distance, $D_{C}(z) = D_{H} \int^{z}_{0} dz' / E(z')$ is the line-of-sight comoving distance, $E(z) = H(z)/H_{0}$, and $\Omega_k \equiv k c^{2}_{0} / H^{2}_{0}$ is the dimensionless curvature density parameter today. One can easily check that the condition for the maximum in $D_{A}$ is now generalized into \cite{MPD87}:
\begin{equation}
\frac{D_{A}(z_{M}) H(z_{M})}{c_{0}} = \begin{cases} \cosh \left( \sqrt{\Omega_{k}} \frac{D_{C}(z)}{D_{H}} \right) &\mbox{for } \Omega_{k} > 0\\
1 &\mbox{for } \Omega_{k} = 0 \\
\cos \left( \sqrt{|\Omega_{k}|} \frac{D_{C}(z)}{D_{H}} \right) &\mbox{for } \Omega_{k} < 0 \, .
\end{cases}
\end{equation}
From the previous expression, we can easily quantify what is the ``error'' in using our Eq.~(\ref{eq:relation}) assuming null curvature. Using the \textit{Planck 2015} data release \texttt{base}$\_$\texttt{omegak}$\_$\texttt{plikHM}$\_$\texttt{TTTEEE}$\_$\texttt{lowTEB}$\_$\texttt{BAO}$\_$\texttt{H070p6}$\_$\texttt{JLA}$\_$\\
\texttt{post}$\_$\texttt{lensing} model, we have $\Omega_k = 0.0008 \pm 0.002$ at the $68\%$ (and $\pm 0.004$ at the $95\%$) confidence level. Assuming for $z_M$ the value of $1.59$ (the maximum for the considered reference model), we obtain a correction $\lesssim0.05 \%$ or, equivalently, the contribution of curvature, in the redshift range of concern, is three-four orders of magnitude less than the leading order which is of interest for us. This result is also in agreement with the recent estimations presented in \citep{BAOK}. Finally, the consistency of the curvature with a null value is generally assumed as an indication of no spatial curvature; this makes us confident about the use of Eq.~(\ref{eq:relation}).

Eq.~(\ref{eq:relation}) itself is very interesting already at this stage: it states that it will be possible to measure  the speed of light \textit{cosmologically}. So far, this has been done only in laboratories on Earth \citep{velocity_c_2}, and such measurements are officially used to establish the value of $c_0$ \citep{velocity_c_1}. Here we have a first ever way to measure the speed of light out of Earth, out of the Solar System, out even of our galaxy. And this measurement is \textit{direct}, imprinted in the clustering of galaxies; as direct as any other measurement that can be done in any terrestrial laboratory.

We also point out another important property of Eq.~(\ref{eq:relation}): it is valid \textit{independently of the cosmological model}, or, in other words, \textit{the measurement of $D_A$ and $H$ at the maximum redshift $z_M$ is unequivocally equal to the value of the speed of light at that time}. 
In Fig.~(\ref{fig:DAH_vs_c}), in the top panels, we plot the quantity $D_A \cdot H$ for different models, chosen from the $10^4$ described above, having different cosmological backgrounds, but assuming a constant speed of light. It is clearly shown that, independently of the value of $z_M$, which is actually dependent on the cosmological background, the quantity $D_A \cdot H$ can change its profile, but when evaluated at $z_M$, is \textit{always} equal to $c_0$. We anticipate here some discussion from the next section, in order to show, in the same Fig.~(\ref{fig:DAH_vs_c}), but in the middle panels,  what happens when the speed of light is varying. In that case, again, we change the cosmological background as above, \textit{and} assume a varying speed of light (more details on how this can be done are in next sections). From left to right, each model has its own maximum, as well as $D_A \cdot H$ asymptotics; and so, given that the speed of light is a function of redshift, its own value for $c(z_M)$. Now, of course $c(z_M)$ will be different for each model, because it is a function of redshift, and $z_M$ changes from one value to another; but the quantity $D_A \cdot H$, evaluated at $z_M$, still has exactly the value equal to $c(z_M)$ which we expect from the theoretical model we used as input.

As a consequence, one could argue that a VSL theory might be constrained directly from the total observations, with no need of any alternative method. Even if this is in theory true, in reality there would be many caveats and conceptual flaws, similar (if not worse) to those one finds when exploring dark energy properties, and which would make impossible to constrain with good accuracy any VSL. Whether theoretically-based or phenomenologically-given, a VSL, faced in this way, would be just an uncertainty (or an ``ignorance'') adding up to other well known uncertainties (``ignorances'') from the cosmological side like, for example, dark energy equation of state or density. Actually, we ignore, in the same way, the ``right'' dark energy behaviour and the ``right'' VSL theory; and they are also degenerate. In fact, VSL were originally introduced as an alternative to inflation (or any accelerated expansion), given that a higher speed of light in the past would solve the horizon event. But a VSL might also mimic dark energy: instead of an energy-mass fluid, dark energy could be explained, totally or in part, as a ``virtual'' effect coming from VSL. Again, we stress that our method is different: we will \textit{measure the speed of light directly}, no indirect inference will be applied.

Thus, our algorithm has to pass through two main steps: \textit{(1.)} the detection of the maximum redshift $z_M$ in $D_A$; \textit{(2.)} the measurement of the speed of light at $z_M$ using $D_A \cdot H$. In the following subsections we will describe in detail the basis for both the steps, highlighting problems and solutions.

\subsection{Maximum detection by BAO}

Eq.~(\ref{eq:relation}), stated in a different form, can be written as
\begin{equation}\label{eq:relation_2}
y_t(z_M) = y_r(z_M)\; ,
\end{equation}
where $y_t$ and $y_r$ are the tangential and radial modes (apart from a multiplicative term equal to the sound horizon which appears both on the right-hand and left-hand side, and having no influence on our results) which will be directly measured by a BAO survey in the next future.

\begin{figure*}[htbp]
\centering
\includegraphics[width=16cm]{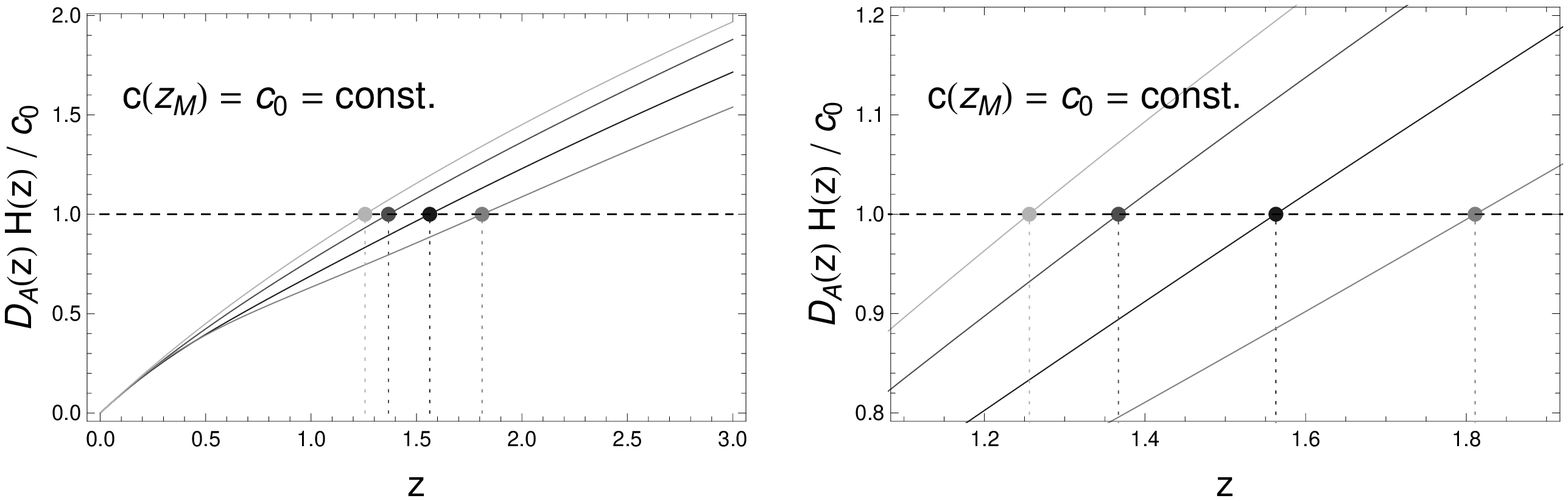}\\
\includegraphics[width=16cm]{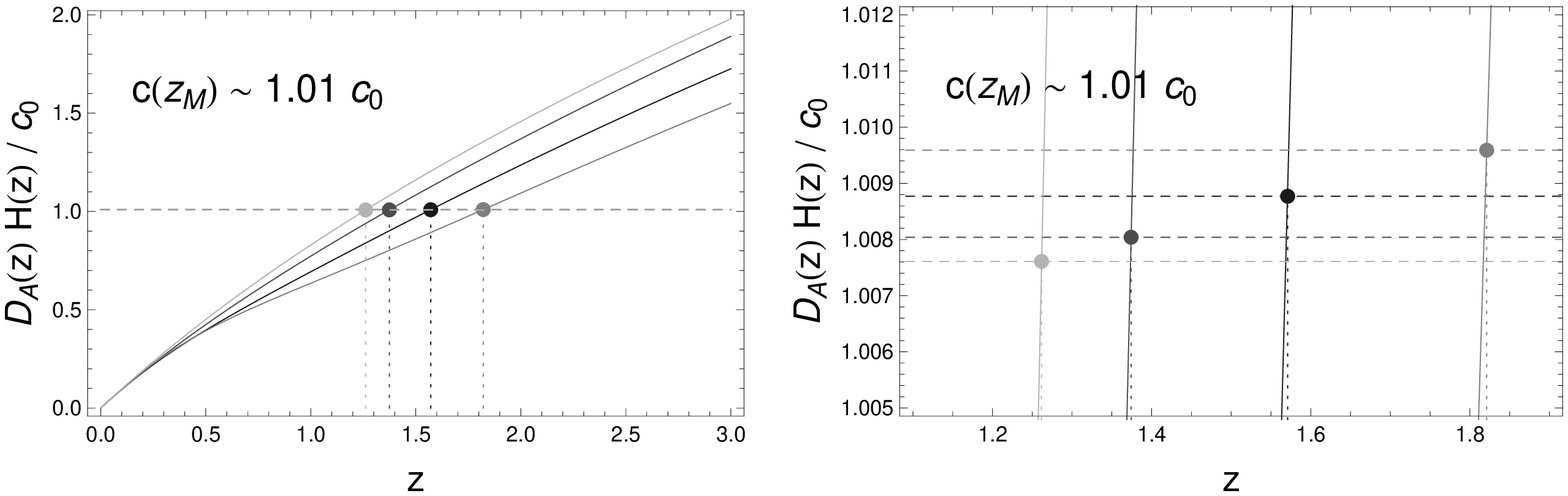}\\
\includegraphics[width=8cm]{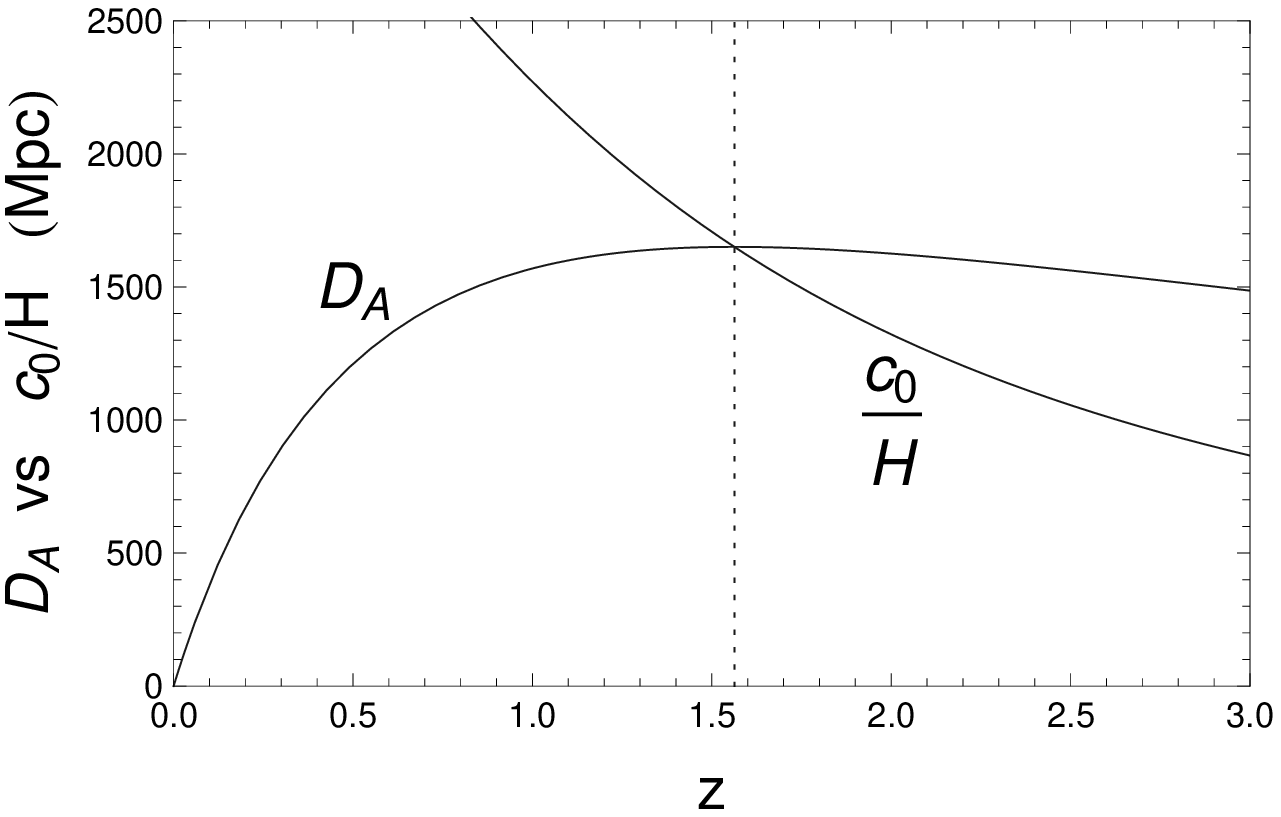}
\caption{\textit{Top panel.} Speed of light measurement through $D_A \cdot H$ evaluated at the maximum redshift $z_M$ when the speed of light is constant. \textit{Middle panel.} Speed of light measurement through $D_A \cdot H$ evaluated at the maximum redshift $z_M$ when the speed of light is a redshift function (see next sections for its formulation): each model recovers its own value for the speed of light.  \textit{Bottom panel.} Determination of the maximum redshift $z_M$.}\label{fig:DAH_vs_c}
\end{figure*}

Eq.~(\ref{eq:relation_2}) is very important for our purposes because it helps to state the determination of the maximum in an easier observationally tested way, equivalent to the vanishing derivative condition, but more precise. In fact, the use of $D_{A}$ only to determine the position of the maximum would be problematic, as a large number of effects combine to smear out the profile of $D_{A}(z)$: the plateau at about $z_{M}$; the measurements of $D_{A}(z)$ from just a few redshift bins from a BAO survey; the errors on the same measurements plus their intrinsic dispersion. The final consequence is the practical impossibility to determine the location of the maximum. But Eq.~(\ref{eq:relation_2}) contains the solution for this problem, at least, when we will have disentangled BAO modes measured by future survey: having at our disposal separate measures of $y_t(z)$ and $y_r(z)$, we can, in principle, constrain the value of $z_{M}$ with better precision, because instead of searching for the maximum in $D_{A}(z)$, one can search for the redshift where the condition $y_t(z_M) = y_r(z_M)$ holds. This is shown at the bottom panel of Fig.~\ref{fig:DAH_vs_c}.

Anyway, even in this case, we can have a better measurement of $z_M$ with respect to the single use of $D_A$, but still it would be far from being useful to measure $c_0$ or some (possible) variation with enough accuracy. But one can employ some cosmologically model-independent method to extract information from data. Literature about this topic is huge, and is growing faster and larger, pulled by the need for disentangling dark energy models in a way which should be as free as possible of theoretical inputs, thus giving independent hints to theory for further developments, confirmations or rebuttal. A non-exhaustive list of such literature is in \citep{Reconstruction_Methods}. We have finally decided to use Gaussian Processes (GPs) \citep{GP,Seikel2012}, which are very well suited to our needs. We postpone a more detailed description of all the details of our implementation of GPs; we just anticipate here that we have employed them to reconstruct $y_t(z)$ and $y_r(z)$ in order to find $z_{M}$. The application of GPs to the BAO modes yields $y_t(z)$ and $y_r(z)$ numerically-reconstructed as smooth analytical functions, which can be evaluated at whatever redshift value one may need; and the sets of GP-reconstructed BAO modes can eventually be employed in a numerical algorithm to estimate $z_{M}$ and its error. Finally, once $z_M$ is known, it will be straightforward to check whether $D_{A}(z_{M}) \cdot H(z_{M}) = c_{0}$ or not.

\subsection{Varying speed of light theories}

In a standard context where the speed of light is not expected to change, combining the errors on $z_{M}$ with the errors on $D_{A}(z) \cdot H(z)$ will measure $c_{0}$ with some error, as it follows from Eq.~(\ref{eq:relation}). Nowadays, the measurement of $c_{0}$ is assumed to be exact and is used as ruler for the definition of the meter \citep{velocity_c_1}. The best measurement for $c_{0}$, obtained with laser interferometry in a terrestrial laboratory, has a relative error $\sim 10^{-9}$ \citep{velocity_c_2}: this precision is largely out of the possibility of a cosmological measurement. But if we assume a VSL, i.e. the existence of a - up to now unknown - function $c(z)$ (with the limit $c(z\rightarrow0) \equiv c_{0}$), then we can recalculate the $\partial D_{A} / \partial z$ in this case, and we would find out that Eq.~(\ref{eq:relation}) would change to the more general expression
\begin{equation}\label{eq:relation_vsl}
D_{A}(z_{M}) \cdot H(z_{M}) = c(z_{M})\, ,
\end{equation}
where, possibly, $c(z_{M}) \neq c_{0}$ is the value of the speed of light at redshift $z_{M}$. Deviations from $c_{0}$, defined from now on through the parameter $\Delta c \equiv c(z_{M})- c_{0}$, if any, can be of whatever order possible, not necessarily as small as $10^{-9}$.

About the approach to follow, we have to advise that, so far, no definitive theoretical background exists for VSL. We have chosen to follow the approach summarized in \citep{Barrow,Magueijo2003}, where a minimal coupling is assumed between matter and the field driving the change in the speed of light. More recent approaches are in \citep{new_Moffat}; but we stress here that, for our needs, the choice of the approach is unimportant.

For our objectives, it is important to check what are the modifications induced by a VSL approach to the Friedmann and continuity equations. In particular, following \citep{Barrow,Magueijo2003}, the first Friedmann equation will look like:
\begin{equation}
H^2(t) = \frac{8\pi G}{3} \rho(t) - \frac{k}{a^2(t)}c^{2}(t)\; ,
\end{equation}
while the continuity equation is:
\begin{equation}
\dot{\rho}(t) + 3 H(t) \left( \rho(t) + \frac{p(t)}{c^2(t)}\right) = \frac{3 k}{4\pi G a^2(t)}c(t)\dot{c}(t)\; ,
\end{equation}
where: $\rho$ and $p$ are, respectively, the energy-mass density and the pressure of any fluid in the Universe; $a(t)$ is the scale factor; $G$ is the universal gravitational constant; and the speed of light is expressed as a general function of time (or redshift), $c(t)$. What is interesting to note, is that any change produced by a VSL is connected with the spatial curvature. Thus, in our case, where we are working assuming the condition of spatial flatness, e.g. $k=0$, this implies that no effective change is working in the continuity equation and, consequently, in the first Friedmann equation which, we underline, is directly connected to the observable quantity $H(z)$.

On the other hand, this is not the only change produced by a VSL; in fact, the speed of light enters all the metric-derived terms like, for example, the expressions for cosmological distances, as $D_A$ is, which involve integrals of the type
\begin{equation}
\int^{z_{2}}_{z_{1}} \frac{c_{0}}{H(z')} \, d \, z' \; ;
\end{equation}
a VSL modifies such integrals in this way:
\begin{equation}
\int^{z_{2}}_{z_{1}} \frac{c(z')}{H(z')} \, d \, z' \; .
\end{equation}

Having clarified the VSL scenario we will work with (but again stressing that we need it only to produce some mock data which include a VSL; thus, the choice of a model over another is not important for our purposes), we have to show now that a general result, independent of the choice of $c(z)$, is that, still, even if we were assuming not negligible spatial curvature, the contribution of $k\neq0$ to our Eq.~(\ref{eq:relation_vsl}) would be many orders smaller than a possible deviation of $c(z_M)$ from $c_0$. In fact, in VSL, Eq.~(\ref{eq:DA_k}) is generalized to:
\begin{equation}\label{eq:DA_k_VSL}
D_{A}(z) = \begin{cases} \frac{D_{H}}{\sqrt{\Omega_{k}}(1+z)} \sinh \left(\sqrt{\Omega_{k}}  \frac{D_{C}(z)}{D_{H}} \right) &\mbox{for } \Omega_{k} > 0\\
\frac{D_{C}(z)}{1+z} &\mbox{for } \Omega_{k} = 0 \\
\frac{D_{H}}{\sqrt{|\Omega_{k}|}(1+z)} \sin \left( \sqrt{|\Omega_{k}|} \frac{D_{C}(z)}{D_{H}} \right) &\mbox{for } \Omega_{k} < 0 \, ,
\end{cases}
\end{equation}
where now the line-of-sight comoving distance is defined as $D_{C}(z) = D_{H} \int^{z}_{0} \Delta_{c}(z') / E(z') dz'$, and we have made use of the general ansatz $c(z) \equiv c_{0} \Delta_{c}(z)$, with $\Delta_{c}(z) = 1$ for $z=0$. From this set of equations, the condition for the maximum of $D_{A}$ within VSL is:
\begin{equation}
\frac{D_{A}(z_{M}) H(z_{M})}{c(z_{M})} = \begin{cases} \cosh \left( \sqrt{\Omega_{k}} \frac{D_{C}(z)}{D_{H}} \right) &\mbox{for } \Omega_{k} > 0\\
1 &\mbox{for } \Omega_{k} = 0 \\
\cos \left( \sqrt{|\Omega_{k}|} \frac{D_{C}(z)}{D_{H}} \right) &\mbox{for } \Omega_{k} < 0 \, .
\end{cases}
\end{equation}
Differently from the case we have considered in the previous section, where the speed of light was constant, here we need to make some assumption on the function form of $c(z)$ in order to quantify the deviation between non-zero curvature assumption and our used formula, Eq.~(\ref{eq:relation_vsl}). If we use for $c(z)$ the expression we will describe in the next section, i.e. our Eq.~(\ref{eq:ansatz_c}); and consider the cases we are going to describe later; and the value of the curvature from \textit{Planck} we have used above, then we can easily find out that even in this case, the error we make in non considering curvature contribution is $\lesssim 0.05 \%$. One important point we have to stress here is that, in principle, some degeneracy could arise between VSL and curvature: the possible detection of a signal might be equally interpreted as ``VSL+null curvature'' or ``constant $c(z)$+curvature''. But this misleading interpretation has a reason to exist only if the VSL signal should result to be of the same order of curvature one, i.e., $\sim 0.01 \%$. Larger detections (if any), could be attributed to VSL only.

\section{Method implementation}
\label{sec:Method}

Once we have defined all the theoretical issues at the base of our test, we can now move on to giving more technical details about how we have built our algorithm and how we checked it to be working well.

\subsection{Mock datasets}

First of all, we have to face one problem: we do not have, now, any data concerning $y_t$ and $y_r$, and neither any BAO observation in the redshift range we need. Thus, we will have to work with mock data. Nowadays, we have no clear and reliable phenomenological expression for $c(z)$ \citep{VSL_observations}; we have chosen to work with a general theoretically-motivated expression given in \citep{Magueijo2003}, i.e.
\begin{equation}\label{eq:ansatz_c}
c(a) \propto c_{0} \left( 1+ a/a_{c} \right)^{n}\; ,
\end{equation}
where, again, $a \equiv 1/(1+z)$ is the scale factor, and $a_{c}$ is the transition epoch from some $c(a) \neq c_{0}$ (at early times) to $c(a) \rightarrow c_{0}$ (at late times - now). Another possible ansatz could be $c \propto c_{0} a^{n}$ \citep{Barrow}, but it resulted to be less flexible in order to (qualitatively) match both early and late times observations, and seems to be inconsistent with experiments \citep{Magueijo2003}. Of course, the choice of the functional form of $c(z)$ is only needed to simulate some mock observational data with some intrinsic variation of $c$, in order to test whether our method is able to detect it or not, and has no influence at all on the final results.

We have decided to produce data based on three different cosmological models:
\begin{itemize}
 \item $\Delta c / c_0 = 0 \%$: the baseline $\Lambda$CDM model from \textit{Planck} 2015 release,  \texttt{base}$\_$\texttt{plikHM}$\_$\texttt{TTTEEE}$\_$\texttt{lowTEB}$\_$\texttt{lensing}$\_$\texttt{post}$\_$\texttt{BAO}. The dimensionless matter density today, inferred by this model, is $\Omega_m = 0.31$;
 \item $\Delta c / c_0 \sim 0.1 \%$\footnote{Note that the effective variation of $c(z)$ at maximum redshift expected from the fiducial model is not exactly equal to $0.1\%$, but slightly less, $\sim 0.08\%$. The same holds true also for the $1\%$ case, which is more exactly $\sim 0.8\%$. We used the $0.1\%$ and $1\%$ notation just for an easier readability.}. at $z_M \approx 1.55-1.6$: baseline $\Lambda$CDM model plus a $c(a)$ given by Eq.~(\ref{eq:ansatz_c}), with $a = 0.05$, $n = -0.001$;
 \item $\Delta c / c_0 \sim 1 \%$ at $z_M \approx 1.55-1.6$: baseline $\Lambda$CDM model plus a $c(a)$ given by Eq.~(\ref{eq:ansatz_c}), with $a = 0.05$, $n = -0.01$.
\end{itemize}
In order to make the global dynamics of the Universe within these two VSL scenarios compatible with present data, we have had to change the value of $\Omega_{m}$, the dimensionless matter density today. This is expected, because, as we have explained in previous sections, VSL can mimic the effects of a dark energy fluid, i.e., an accelerated expansion. A higher speed of light in the past can mimic the effects of a dark energy component, thus resulting in a lower value for $\Omega_{DE}$ (dimensionless dark energy density today). Equivalently, when no spatial curvature is assumed, this gets converted to a larger value of $\Omega_{m}$. In order to arrange for the above assessed variations in $c$, in the second model of VSL, we need $\Omega_{m} = 0.314$, and in the third we need $\Omega_{m} = 0.348$, while the value for the first case (thus, constant $c_0$) is $\Omega_m = 0.31$. We stress again that such values are not derived from a fitting procedure to present cosmological data, which is out of the purpose of this work. We simply checked heuristically the values which could give a qualitatively good global description of present data. As a proof of such goodness, in Table~\ref{tab:model_vs_data} we calculate all the quantities of interest for all the three models we have considered, and we compare them with the available measurements. The sound horizon at decoupling, $r_s(z_{\ast})$ is derived from the same baseline model from \textit{Planck 2015} used to mimic data as described above; the BOSS Data Release 11 data are from \citep{BOSS}; the WiggleZ Dark Energy Survey data are from \citep{WiggleZ}; while the $H(z)$ data from cosmic chronometers are from \citep{chronometers_2}. We can easily check that the changes in the sound horizon are $\lesssim \sigma_{s}$ in all the cases considered, where $\sigma_{s}$ is the error from the chosen \textit{Planck} fiducial model. The same holds true for the angular diameter distances measurements, which are all consistent with data in the error confidence level, $\sigma_{D_{A}}$, we have from present surveys. For the rate expansion $H$ we have some more tension with data (bold numbers); but, still, our proposed VSL models are consistent with the theoretical $\Lambda$CDM model from \textit{Planck} results which we use as fiducial in a ``standard context''.

{\renewcommand{\tabcolsep}{1.5mm}
{\renewcommand{\arraystretch}{1.5}
\begin{table}[htbp]
\begin{minipage}{0.5\textwidth}
\caption{Qualitative comparison among data and models. The distances (sound horizon at decoupling, $r_s(z_{\ast})$; angular diameter distances, $D_A$) are in $Mpc$; rate expansion data (Hubble function, $H$) are in $km$ $s^{-1}$ $Mpc^{-1}$.}\label{tab:model_vs_data}
\centering
\resizebox*{0.99\textwidth}{!}{
\begin{tabular}{|c|c|c|c|c|}
\hline \hline
 & Data & $\Lambda$CDM & $\Delta c/c_0 = 0.1\%$ & $\Delta c/c_0 =1\%$ \\
\hline
\hline
 \multicolumn{5}{|c|}{\textit{Planck 2015}}\\
\hline
 $r_{s}(z_{\ast})$ & $144.77 \pm 0.24$ & $144.70$ & $144.67$ & $144.75$ \\
\hline
\hline
 \multicolumn{5}{|c|}{\textit{BOSS}}\\
\hline
 $D_A(z=0.57)$ & $1380 \pm 23 $ & $1388$ & $1385$ & $1371$ \\
 $H(z=0.57)$ & $93.1 \pm 3.0$ & $93.0$ & $93.3$ & $95.6$ \\
 $D_A(z=2.34)$ & $1662 \pm 96$ & $1730$ & $1725$ & $1684$ \\
 $H(z=2.34)$ & $222 \pm 7$ & $\mathbf{237}$ & $\mathbf{238}$ & $\mathbf{250}$ \\
\hline \hline
 \multicolumn{5}{|c|}{\textit{WiggleZ}}\\
\hline
 $D_A(z=0.44)$ & $1204.9 \pm 113.6$ & $1196.2$ & $1208.5$ & $1198.0$\\
 $H(z=0.44)$ & $82.6 \pm 7.8$ & $88.0$ & $86.2$ & $88.0$ \\
 $D_A(z=0.60)$ & $1380.1 \pm 94.8$ & $1400.5$ & $1419.2$ & $1403.4$ \\
 $H(z=0.60)$ & $87.9 \pm 6.1$ & $\mathbf{97.5}$ & $\mathbf{95.0}$ & $\mathbf{97.5}$ \\
 $D_A(z=0.73)$ & $1533.7 \pm 106.8$ & $1517.0$ & $1540.6$ & $1520.6$\\
 $H(z=0.73)$ & $97.3 \pm 7.0$ & $\mathbf{106.0}$ & $102.9$ & $\mathbf{106.0}$ \\
\hline \hline
 \multicolumn{5}{|c|}{\textit{Cosmic Chronometers}}\\
\hline
 $H(z=0.070)$ & $69.0 \pm 19.6$ & $69.97$ & $70.00$ & $70.25$ \\
 $H(z=0.090)$ & $69 \pm 12$     & $70.68$ & $70.72$ & $71.04$ \\
 $H(z=0.120)$ & $68.6 \pm 26.2$ & $71.77$ & $71.82$ & $72.26$ \\
 $H(z=0.170)$ & $83 \pm 8$      & $\mathbf{73.69}$ & $\mathbf{73.76}$ & $\mathbf{74.40}$ \\
 $H(z=0.179)$ & $75 \pm 4$      & $74.05$ & $74.13$ & $75.79$ \\
 $H(z=0.199)$ & $75 \pm 5$      & $74.86$ & $74.94$ & $75.69$ \\
 $H(z=0.200)$ & $72.9 \pm 29.6$ & $74.90$ & $74.99$ & $75.74$ \\
 $H(z=0.270)$ & $77 \pm 14$     & $77.87$ & $78.00$ & $79.03$ \\
 $H(z=0.280)$ & $88.8 \pm 36.6$ & $78.32$ & $78.44$ & $79.52$ \\
 $H(z=0.352)$ & $83 \pm 14$     & $81.63$ & $82.80$ & $83.19$ \\
 $H(z=0.400)$ & $95 \pm 17$     & $83.97$ & $84.16$ & $85.76$ \\
 $H(z=0.480)$ & $97 \pm 62$     & $88.08$ & $88.31$ & $90.26$ \\
 $H(z=0.593)$ & $104 \pm 13$    & $94.30$ & $94.60$ & $97.07$ \\
 $H(z=0.680)$ & $92 \pm 8$      & $99.42$ & $99.77$ & $\mathbf{102.64}$ \\
 $H(z=0.781)$ & $105 \pm 12$    & $105.70$ & $106.10$ & $109.46$ \\
 $H(z=0.875)$ & $125 \pm 17$    & $111.85$ & $112.31$ & $116.12$ \\
 $H(z=0.880)$ & $90 \pm 40$     & $112.18$ & $112.64$ & $116.48$ \\
 $H(z=1.037)$ & $154 \pm 20$    & $\mathbf{123.08}$ & $\mathbf{123.64}$ & $\mathbf{128.24}$ \\
 $H(z=1.300)$ & $168 \pm 17$    & $\mathbf{142.90}$ & $\mathbf{143.61}$ & $\mathbf{149.54}$ \\
 $H(z=1.363)$ & $160.0 \pm 33.6$ & $147.91$ & $148.66$ & $154.91$ \\
 $H(z=1.430)$ & $177 \pm 18$    & $\mathbf{153.35}$ & $\mathbf{154.14}$ & $160.74$ \\
 $H(z=1.530)$ & $140 \pm 14$    & $\mathbf{161.66}$ & $\mathbf{162.52}$ & $\mathbf{169.64}$ \\
 $H(z=1.750)$ & $202 \pm 40$    & $180.73$ & $181.73$ & $190.02$ \\
 $H(z=1.965)$ & $186.5 \pm 50.4$ & $200.35$ & $201.49$ & $210.95$ \\
\hline \hline
\end{tabular}}
\end{minipage}
\end{table}}}

In \citep{FontRibera2014} many on-going and future surveys are analyzed, among them: BOSS, eBOSS, HETDEX, DESI, \textit{Euclid} and \textit{WFIRST-2.4}. In particular, the authors focus on the constraints on $D_{A}$ and $H$ from BAO analysis and conclude that the best results are from the ESA mission \textit{Euclid}: in Table~6 of \citep{FontRibera2014} they show the percentage errors on $D_{A}/r_s(z_{\ast})$ and $H \cdot r_s(z_{\ast})$ for $15$ redshift bins (of width $0.1$) in the redshift range $[0.6;2.1]$  covered by \textit{Euclid}. Once we have the fiducial mock data, $D^{fid}_{A}/r^{fid}_s(z_{\ast})$ and $H^{fid} \cdot r^{fid}_s(z_{\ast})$, derived from the three models described above, we can easily calculate the corresponding errors $\sigma_{D_{A}/r_s(z_{\ast})}$ and $\sigma_{H \cdot r_s(z_{\ast})}$ from the columns $2$ and $3$ in Table~6 of \citep{FontRibera2014}.

\textit{Euclid} will be considered like a sort of ``pessimistic'' scenario in our work, because, at least using the available forecast estimation we have now, the Square Kilometer Array (SKA) \citep{SKA} should be much better than \textit{Euclid}, even if in a smaller redshift range, but still reaching the values we need in order to determine the maximum redshift $z_M$ (thus, up to $z \approx 1.8$). In \citep{SKA} the percentage errors on $D_A$ and $H$ expected from this survey are shown in their Fig.~5; we can use them, once given the fiducial values for these quantities, to calculate their corresponding errors. In our work, results from SKA will be an ``optimistic'' scenario.

Anyway, we will not work directly on the fiducial model values. Instead, we will randomly pick up values of $D_{A}/r_s(z_{\ast})$ and $H \cdot r_s(z_{\ast})$ (or $D_A$ and $H$) from a multivariate Gaussian centered on the fiducial values, and with a total covariance matrix built up from the errors we derived in the way previously described, and assuming an additional correlation factor between them, equal to $r \sim 0.4$, as derived in \citep{SeoEisenstein2007}. Such procedure is needed in order to give to mock data an intrinsic dispersion closer to the real one. Finally, of course, we cannot rely on the results from only one single random run. Instead, we produce $10^3$ random mock data sets, in the way just described, and we test our algorithm on each of them. Thus, our final results will be then a statistical output on an ensemble of possible universes observationally compatible with the starting fiducial model. We want to clarify that this (i.e., the making of mock cosmological data) is the only step in our work where we need to assume a cosmological model. This choice is quite unavoidable in order to have a reference point to establish the goodness of our analysis, but it is a quite common procedure in forecast analysis. Moreover, the choice to test our method on a large number of data sets will greatly smear the effects of this initial input which, anyway, it is absolutely not in contrast with the (up to some limit, as clarified in the previous section) model-independence of our method.

\subsection{Gaussian Processes}

Once we have our set of data and related errors, we can apply GPs in order to reconstruct the observational quantities of interest ($y_{r}$ and $y_{t}$) and calculate the position of the maximum $(z_M)$. GPs are very helpful because they incorporate in a very natural and straightforward way correlations between data, even when expressed in the form of a non-diagonal covariance matrix, which is our case now. For all the details about GPs see the related literature \citep{GP,Seikel2012}; here, we will discuss in more detail only some aspects of their implementation, necessary for our purposes:
\begin{itemize}
  \item in \citep{VSL_us} we used a simple Gaussian as the covariance function relating two points at different redshifts, $z$ and $\bar{z}$. In \citep{Seikel2012} it is shown that such choice, for the quantities we are considering, can lead to an underestimation of the errors of the reconstruction. A more suitable choice, in this sense, would be the Mat$\mathrm{\acute{e}}$rn$(9/2)$ function, given by
      \begin{eqnarray}
      k(z,\bar{z}) &=& \sigma^{2}_{f} \exp\left[-\frac{3(z-\bar{z})}{l}\right] \nonumber \\
      &\times& \left(1+ \frac{3|z-\bar{z}|}{l}+\frac{27(z-\bar{z})^2}{7l^2} \right. \nonumber \\
      &+& \left. \frac{18|z-\bar{z}|^3}{7l^3} + \frac{27(z-\bar{z})^4}{35l^4} \right)\; ,
      \end{eqnarray}
      where $\sigma_f$, the signal variance, and $l$, the characteristic length scale, are the hyperparameters of the proposed correlation;
  \item for each one of the $10^3$ mock data set we have created, we employ a Markov Chain Monte Carlo Method in order to find the values of the hyperparameters which optimize the reconstruction of $D_A$ and $H$, following \citep{Seikel2012};
  \item once we have found such optimized reconstruction parameters, we evaluate the GPs output functions, i.e. $y_{t}$ and $y_r$, on a $\Delta z = 0.01$ redshift grid, ten times finer than the \textit{Euclid} and SKA forecasted bins;
  \item such a finer grid is useful to implement a numerical algorithm to calculate $z_{M}$ for each simulation. We finally have $10^3$ sets of GPs-reconstructed $(y^{GP}_t,\sigma^{GP}_{y_t})$ and $(y^{GP}_r,\sigma^{GP}_{y_r})$ and, using our Eq.~(\ref{eq:relation_2}), we can estimate $z_M$ for each of them.
\end{itemize}
Such a finer grid is useful to implement a numerical algorithm to calculate $z_{M}$ for each simulation:
\begin{itemize}
  \item we have $10^3$ sets of GPs-reconstructed $(y^{GP}_t,\sigma^{GP}_{y_t})$ and $(y^{GP}_r,\sigma^{GP}_{y_r})$: we randomly pick up $\sim 160$ sets in the $[-4\sigma^{GP},4\sigma^{GP}]$ confidence level for each quantity, that we then combine obtaining a total of $\sim 2.5 \cdot 10^4$ ($y^{GP}_t, y^{GP}_r$) pairs;
  \item for each pair, we fit $y^{GP}_t$ and $y^{GP}_r$ with a high order polynomial in the redshift range $[1.,2.]$ (for \textit{Euclid}; the maximum redshift is $1.8$ for SKA) and we find $z_{M}$ numerically for each of them using Eq.~(\ref{eq:relation_2}).
\end{itemize}
Thus we end with a set of $\sim 2.5 \cdot 10^4$ $z_{M}$ from which we can derive the mean value $z_{M}$ for our statistical ensemble, and the related error $\sigma_{z_{M}}$.

\subsection{Speed of light measured}

Once we have $(z_M, \sigma_{z_M})$, we only need to calculate the quantity $D_{A}(z) \cdot H(z)/c_{0}$, using the GPs-reconstructed data set and, using our Eqs.~(\ref{eq:relation}) and (\ref{eq:relation_vsl}), we can constrain the value of the speed of light. We choose to normalize the quantity $D_{A}(z) \cdot H(z)$ with $c_{0}$; thus, in the context of constant speed of light, we expect to find $D_{A}(z_{M}) \cdot H(z_{M})/c_{0} \approx 1$ with some error, while in VSL theories it can be $\neq 1$.

One important question deserves to be discussed at this point: the very useful relations to determine the speed of light at far cosmological epochs is Eq.~(\ref{eq:relation}); Eq.~(\ref{eq:relation_2}) is a completely equivalent way to write it, absolutely needed for the determination of $z_{M}$, but quite useless for checking the constancy of the speed of light. Stated in another way: BAO are necessary for the determination of $z_M$, using Eq.~(\ref{eq:relation_2}), but cannot be used to measure the speed of light, using Eq.~(\ref{eq:relation}). This is clearly understandable if we take a careful look at the way the radial mode, which can be measured from BAO, is defined: the measured length, $y_r$, which is actually what one can see in the galaxy distribution, combines information from both the speed of light and the expansion rate $H$. In order to use our Eq.~(\ref{eq:relation}), we need to determine $H$ from what we see, but we cannot actually use the BAO radial model $y_r$, because in this case we would need some assumption on the speed of light functional form, which is, of course, unknown to us (at least, if one assumes it can be varying).

But even if we cannot use BAO from SKA, \textit{Euclid} or \textit{WFIRST-2.4}, there is still a way these surveys (at least, the optical ones, i.e. \textit{Euclid} or \textit{WFIRST-2.4}) can be useful to us: as said in the Introduction, the same galaxies which are used to measure BAO (or, better, the fraction of them corresponding to ETGs), can be used as cosmic chronometers, thus giving us direct measurements of $H(z)$, free of any degeneracy and/or assumption on the possible time variability of $c(z)$. Such use is quite interesting, because we can note down the close analogy of such probes with a laboratory experiment: here we would have $D_{A}$ from BAO, a length, which plays the role of a standard (cosmological) ruler; and $H^{-1}$ from cosmic chronometers, with the dimension of time, as a (cosmological) clock.

A preliminary study about the capability of future surveys in this context is discussed in \citep{chronometers_2}, where a simulated \textit{Euclid}-type survey gives a minimum $\sim 5\%$ error on $H$, when accounting for statistical errors only. Better performances should be obtained with \textit{WFIRST-2.4}, which is going to observe more galaxies than \textit{Euclid}. However, in the following, we will assume that the errors on $H$ are those expected from BAO observations, which are smaller than this limit. This does not invalidate our method; in some way our analysis will give us a clear idea about how much can we expect from it, and will give us an indication about the properties a survey should have in order to be sensitive to the signal we are searching for. If an already-planned survey is not going to reach such limit, this does not exclude the possibility that, in the future, we can have such a detection by means of more advanced instruments.

\section{Results and discussion.}
\label{sec:Results}

\begin{figure*}[htbp]
\includegraphics[width=15cm]{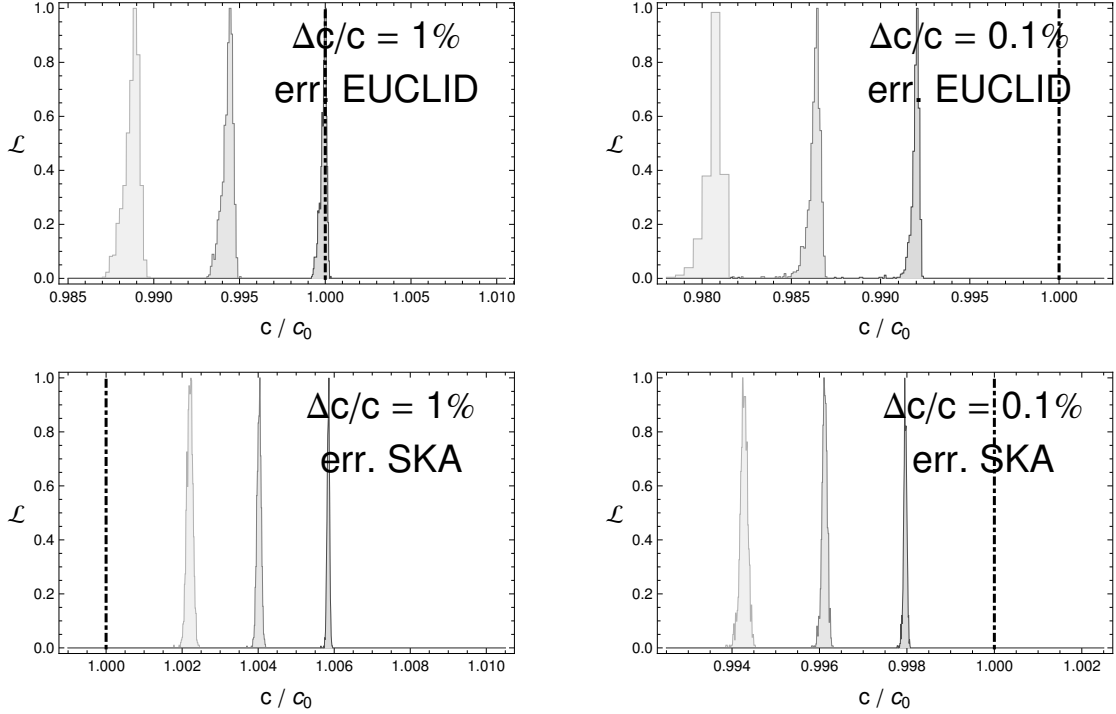}
\caption{Probability distribution of $c_{1\sigma}$, $c_{2\sigma}$ and $c_{3\sigma}$ (from dark to light grey) from $10^3$ simulations in different survey configurations. Vertical black dot-dashed line is for $c(z_{M}) = c_{0}$.}\label{fig:histograms}
\end{figure*}

All the results from applying our method to the mock data we have produced, are summarized in Table~\ref{tab:results} and visualized in Fig.~(\ref{fig:histograms}), where we show, separately, expectations from assuming errors on $D_{A}$ and $H$ as they come from \textit{Euclid} and SKA for the three cosmological scenarios we have considered. Note that the errors shown in the table are not the usual ones, except for those on $z_M$. Instead, given that we have been working on an ensemble of $10^3$ possible observational sets, we give our results in the form ``average of the median from the ensemble $\pm$ average of the median of the standard deviation from the ensemble''. We also have to specify the notation we used: $c$ is the measured value from the experiment, i.e., $c(z_M)$, normalized to the present value $c_0$; $c_{1\sigma}$, $c_{2\sigma}$, $c_{3\sigma}$, are the lower limits at, respectively, $1\sigma$, $2\sigma$ and $3\sigma$ confidence levels. We only consider the lower limits, because in all the models we have assumed, the speed of light in the past was greater than the present value; thus, any deviation from constancy is possibly detectable only if the lower limits are greater than $c_0$. The choice of models with a different trend would have been completely equivalent, just implying we should have focused on the upper limits instead of the lower ones; but all the conclusions we drive out would have been completely equivalent to the present ones. The $p>1$ number is the probability to have a $c(z_M) \neq c_0$ in our ensemble (e.g., the normalized number of simulations for which a clear non-constant signal can be detected); higher values of $p>1$ mean, of course, that the survey is more likely to observe a deviation from constancy of the speed of light.

As a preliminary check, we have tested our algorithm in the case of non-varying speed of light. The expected maximum from the fiducial model is $z_{M} = 1.589$, and we recover, in the case of SKA, a value that is highly consistent with this estimation. As expected, even the value of $c(z_M)$ is very consistent with the expected $c_0$, and there is a very small dispersion of the values from the $10^3$ models we have considered. Thus, we can finally assess that our method works quite well, as we are able to recover the input model with a very good accuracy.

The central point here is the lower limit detection: from the $\Delta c = 0$ case, we can see that the average $1\sigma$ limit is $\approx 0.003$. So, the main question to be answered now is: \textit{is this accuracy enough to detect a possible VSL?}

To answer, we consider the VSL model with a $1\%$ variation in $c$. First of all, again, the detection of $z_M$ is good: the expected value is $z_{M}= 1.561$, and we recover $1.559$ and $1.561$, respectively, with \textit{Euclid} and SKA, with both the errors fully compatible with the expected input. What is interesting to note is the improvement in the determination of the maximum which is achieved when moving from one survey to another. In particular, the error on $z_M$ from SKA is $\sim 30\%$ smaller than the same estimation from \textit{Euclid}. This, of course, will also result in a better constrain on $c(z_M)$. Effectively, if we examine the $1\sigma$ lower limit, we can see how \textit{Euclid} will be hardly able to detect such order of variation in the speed of light, with only $32\%$ of our simulations clearly detecting a deviation from $1$ (i.e. from $c_0$ and, thus, constancy) at $1\sigma$. On the other hand, SKA will be extremely useful, with a clear detection at even $3\sigma$. We point out that in \citep{VSL_us} we concluded that \textit{Euclid} would have been able to detect such a variation at $2\sigma$; here the signal is worsened by the change of the correlation function in the GPs. Using the Mat$\mathrm{\acute{e}}$rn$(9/2)$ function, instead of a Gaussian, as GPs correlation kernel, made the errors much more realistic, but also larger than those we obtained in \citep{VSL_us}, and this reflects in such new results for \textit{Euclid}.

Things go a little worse for smaller variations: it is clear that a $0.1\%$ variation in $c$ will be very hardly detectable even with SKA: even if the corresponding values for $z_M$ and $c(z_M)$ can be recovered, the sensitivity will not be enough to discriminate between such small deviation and the constancy. Thus, we have to assume that the sensitivity of at least these two already-planned surveys will not be enough to detect a VSL too much smaller than $\sim 1\%$.

For this reason, we have explored whether there is any chance for some future more extreme galaxy survey to perform better. Building a reliable galaxy survey in all possible details has many constructive difficulties and it is out of the purpose of this work. We have thus carried out a naive ``rule-of-thumb'' analysis: we have assumed a SKA-style survey (i.e., with the same redshift range and bins as SKA), but with a better performance, quantified as smaller errors on $D_{A}$ and $H$, and actually possible if the number of observed galaxies is increased. We have first considered the case where the errors on $D_{A}$ and $H$ are reduced by one third (this is approximately the same improvement one has when moving from \textit{Euclid} to SKA). But even in this case, the $0.1\%$ variation is still out of possibility. In order to find out something at the $1\sigma$ level, you need to reduce the errors by a factor of $10$.

It is clear that to put a limit detection to $0.1\%$, for a VSL might be problematic. We are not aware of any cosmological \textit{direct or indirect} measurement of $c$ which can be used as a comparison tool; in most cases, the speed of light is simply assumed to be constant and equal to the common value $c_0$. On the other hand, in the literature, there are many measurements of another quantity which is strictly related to $c$, namely, the fine-structure constant $\alpha$, which is exactly defined as $\alpha \equiv e^{2}/ (\hbar \, c)$, where $e$ is the electron charge and $\hbar$ the reduced Planck constant. There are many observations which are compatible with a varying $\alpha$ \citep{Uzan2011,alpha}, but these variations are always very small, at least $<10^{-4}$; that is the reason why there is debate about whether they are really consistent or, instead, we should more correctly assume that $\alpha$ is constant. However, if we center on its definition, it is easy to check that, if the other parameters involved in its definition $(e, \hbar)$ are assumed to be constant, then $\Delta \alpha / \alpha = - \Delta c / c_{0}$. Thus, we would expect a variation for $c$ of the same-order, i.e $<10^{-4}$. The question here is more subtle, anyway. In fact, in principle, even a large variation in the speed of light could be compatible with such orders of magnitudes for the variation in $\alpha$, if we admit that also the other parameters can vary. But in this case we would have an unpleasant ``fine-tuning'' and degenerate conspiracy plot from many different aspects of physics, because in order to accommodate such small variations in $\alpha$, we would need: \textit{either} a larger variation from each of the other parameters to compensate each other, \textit{or} roughly the same order of magnitude variation for each one of them.

Thus, assuming such measurements of variation of $\alpha$ are correct, it is a conservative assumption to expect the same order of variation for $c$. Reading the literature, it is easy to check that the most used varying $\alpha$ detectors are quasars, in particular, their spectra, and quasars are not \textit{cosmological scale} structures. What we are effectively measuring is a possible \textit{local} variation of $\alpha$. Instead, in our method, we are going to measure $c$ from the \textit{cosmological-scale} distribution of galaxies, which is many order of magnitudes larger than the quasars scale. The only similar probe which might be compared to our results, is the Cosmic Microwave Background (CMB): from the first \textit{Planck} release, a possible limit has been detected on the variation of $\alpha$ at a redshift $z \sim 10^3$ of the order of $\Delta \alpha / \alpha \lesssim 0.4 \%$, see \citep{Planck_26} and \citep{OBrian}. But the constraint from \citep{Planck_26} is plagued by a strong degeneracy with a cosmological parameter, the Hubble constant $H_0$, resulting in a tension which is only mildly reduced by joining CMB observations with some archival BAO data sets, and adding a prior on $H_{0}$. Moreover, it is worth to point out that this is not a \textit{direct} measurement of $\alpha$: a variation is assumed, and then parameterized in a very simple, but arbitrary, way. From this point of view, we would like to stress that \textit{ our ``optimistic-scenario'' of VSL detection from SKA is highly competitive with CMB observations, and it would be obtained without any assumption on other possible cosmological parameters. It would be a direct measurement, and not indirectly inferred.} Then, the constraint from \citep{OBrian} searches for an even more extreme variation: not only time but also spatial variation, for which the signal can be even smaller. And, actually, their conclusion is that even if there is any variation, this is consistent with zero.

Finally, taking into account all such results and arguments, if we focus on our hypothesis about the sensitivity of futuristic surveys, one question which should be answered lastly is: \textit{is it technically possible to achieve such small errors from galaxy surveys and thus be able to measure finer variations of $c$?} From a quantitative point of view, the answer is not easy because, as we have said above, it would involve many technical problems. But, at least qualitatively, we feel enough confident that the $0.1\%$ limit in VSL detection is within the reach of future observations. In \citep{EUCLID_1} (Fig.~2.21) and in \citep{FontRibera2014} (Fig.~3), the observational errors from many on-going and planned future surveys are shown. As it can be easily checked by simple visual inspection, \textit{Euclid} errors are expected to be about one tenth of the errors obtained from an already completed survey like the WiggleZ Dark Energy Survey. Thus this level of improvement we have considered should be possible. Moreover, it is also clear that any technical improvement will be useful: based on modern technology, the ground based telescopes like DESI and SKA \citep{SKA} are already almost as competitive as space ones like \textit{Euclid} and \textit{WFIRST-2.4}, with SKA errors that should be one third of \textit{Euclid} ones. Thus we expect that in the future, even if still not planned, it will be surely possible to further improve space-based surveys and obtain even better constraints. One point we have to remember, however, is that the $H$ measurements from such future galaxy surveys, are strictly related to cosmic chronometers, whose errors are somewhat larger than the ones we have used here, and which were estimated from a BAO analysis. This makes things more difficult, but not impossible.

{\renewcommand{\tabcolsep}{1.5mm}
{\renewcommand{\arraystretch}{2.}
\begin{table*}[htbp]
\begin{minipage}{0.95\textwidth}
\caption{Results.}\label{tab:results}
\centering
\resizebox*{\textwidth}{!}{
\begin{tabular}{|c|ccccc|}
\hline \hline
 & \multicolumn{5}{c|}{\textit{Euclid}} \\
$\Delta c / c_{0}$ & $z_{M}$ & $c \; (p_{>1})$ & $c_{1\sigma}\; (p_{>1})$ & $c_{2\sigma}\; (p_{>1})$ & $c _{3\sigma}\; (p_{>1})$ \\
\hline
$1\%$ & $1.559^{+0.054}_{-0.051}$ & $1.00872^{+0.00003}_{-0.00003} \; (1)$ & $0.99993^{+0.00013}_{-0.00024} \; (0.32)$ & $0.99436^{+0.00023}_{-0.00041} \; (0)$ & $0.98879^{+0.00032}_{-0.00056}\; (0)$ \\
$0.1\%$ & $1.587^{+0.058}_{-0.052}$ & $1.000880^{+0.000006}_{-0.000006} \; (0.98)$ & $0.99199^{+0.00014}_{-0.00024} \; (0.001)$ & $0.98636^{+0.00024}_{-0.00038} \; (0)$ & $0.98072^{+0.00034}_{-0.00053} \; (0)$ \\
\hline \hline
 & \multicolumn{5}{c|}{SKA} \\
$\Delta c / c_{0}$ & $z_{M}$ & $c \; (p_{>1})$ & $c_{1\sigma}\; (p_{>1})$ & $c_{2\sigma}\; (p_{>1})$ & $c _{3\sigma}\; (p_{>1})$ \\
\hline
$0\%$ & $1.593^{+0.018}_{-0.017}$ & $1.^{+3 \cdot 10^{-7}}_{-4 \cdot 10^{-7}}$ & $0.99708^{+0.00003}_{-0.00004}$ & $0.99524^{+0.00006}_{-0.00007}$ & $0.99339^{+0.00008}_{-0.00008}$ \\
$1\%$ & $1.561^{+0.017}_{-0.017}$ & $1.00873^{+0.00001}_{-0.00001} \; (1)$ & $1.00585^{+0.00003}_{-0.00003} \; (1)$ & $1.004036^{+0.00005}_{-0.00005} \; (1)$ & $1.00221^{+0.00008}_{-0.00009}\; (1)$ \\
$0.1\%$ & $1.590^{+0.018}_{-0.017}$ & $1.000880^{+0.000001}_{-0.000001} \; (1)$ & $0.99797^{+0.00003}_{-0.00003} \; (0)$ & $0.99612^{+0.00006}_{-0.00006} \; (0)$ & $0.99428^{+0.00008}_{-0.00008}\; (0)$ \\
$0.1\% \; (err/3)$ & $1.590^{+0.006}_{-0.006}$ & $1.0008800^{+0.0000001}_{-0.0000001} \; (1)$ & $0.999834^{+0.000009}_{-0.000009} \; (0)$ & $0.99917^{+0.00001}_{-0.00001} \; (0)$ & $0.998510^{+0.00002}_{-0.00002}\; (0)$ \\
$0.1\% \; (err/10)$ & $1.590^{+0.003}_{-0.003}$ & $1.0008800^{+0.0000003}_{-0.0000002} \; (1)$ & $1.00032^{+0.00014}_{-0.00018} \; (0.94)$ & $0.99996^{+0.00023}_{-0.00029} \; (0.44)$ & $0.99961^{+0.00032}_{-0.00040}\; (0.10)$ \\
\hline \hline
\end{tabular}}
\end{minipage}
\end{table*}}}

\section{Acknowledgements}

The research of V.S. and M.P.D. was financed by the Polish National Science Center Grant DEC-2012/06/A/ST2/00395. R.L. was supported by the Spanish Ministry of Economy and Competitiveness through research projects FIS2014-57956-P (comprising FEDER funds) and Consolider EPI CSD2010-00064.

\vfill
\end{document}